\documentclass[12pt,preprint,epsf]{aastex}

\newcommand{\ubvri}{\protect\hbox{$U\!BV\!RI$} }
\newcommand{\bvri}{\protect\hbox{$BV\!RI$} }

\shorttitle{Sky Brightness at Cerro Tololo}
\shortauthors{Krisciunas et al.} 

\begin{document}
\received{7 May 2007}

\title{Optical Sky Brightness at Cerro Tololo Inter-American Observatory from 1992 to 
2006\altaffilmark{1}}
\author{
Kevin Krisciunas,\altaffilmark{2}
Dylan R. Semler,\altaffilmark{3}
Joseph Richards,\altaffilmark{4}
Hugo E. Schwarz,\altaffilmark{5}
Nicholas B. Suntzeff,\altaffilmark{2}
Sergio Vera,\altaffilmark{6}
and Pedro Sanhueza\altaffilmark{7}
}
\altaffiltext{1}{Based in part on observations taken at the Cerro Tololo
Inter-American Observatory, National Optical Astronomy Observatory, 
which is operated by the Association of Universities for Research in 
Astronomy, Inc. (AURA) under cooperative agreement with the National 
Science Foundation.}
\altaffiltext{2}{Texas A. \& M. University, Department of Physics,
  4242 TAMU, College Station, TX 77843; 
  {krisciunas@physics.tamu.edu}, {suntzeff@physics.tamu.edu} }
\altaffiltext{3}{Columbia University, 1027 Pupin Hall MC 5247,
550 W. 120th Street, New York, NY 10027; {dsemler@astro.columbia.edu} }
\altaffiltext{4}{Department of Statistics, Baker Hall,
Carnegie Mellon University, Pittsburgh, PA 15213;   {joeyrichar@gmail.com} }
\altaffiltext{5}{Deceased, 20 October 2006.}
\altaffiltext{6}{Cerro Tololo Inter-American Observatory, Casilla 603,
  La Serena, Chile }
\altaffiltext{7}{Oficina de Protecci\'{o}n de la Calidad del Cielo del Norte de Chile (OPCC),
1606 Cisternas, La Serena, Chile; {psanhueza@opcc.cl} }

\begin{abstract} 

We present optical (\ubvri\hspace{-1.0mm}) sky brightness measures from 1992 through
2006.  The data are based on CCD imagery obtained with the CTIO 0.9-m, 1.3-m, and 1.5-m
telescopes.  The $B$- and $V$-band data are in reasonable agreement with 
measurements previously made at Mauna Kea, though on the basis of a small number of
images per year there are discrepancies for the years 1992 through 1994.  Our
CCD-based data are not significantly different than values obtained at Cerro Paranal.
We find that the yearly averages of $V$-band sky brightness are best correlated with
the 10.7-cm solar flux taken 5 days prior to the sky brightness measures.  This
implies an average speed of 350 km s$^{-1}$ for the solar wind.
While we can measure an enhancement of the night sky levels over La Serena 10 degrees
above the horizon, at elevation angles above 45 degrees we find no evidence that
the night sky brightness at Cerro Tololo is affected by artificial light of nearby
towns and cities.

\end{abstract}


\section{Introduction}

A knowledge of the sky background is fundamental to optical and infrared observational
astronomy.  The accuracy of photometric measurements hinges on the signal-to-noise {\em
ratio}, so we would like the noise to be as small as possible. The noise has a number
of components, among them the dark counts (or dark current), the readout noise, the sky
background, and the cosmic ray flux. Furthermore, an instrument must be matched to the
typical seeing at a given site, and that stipulates an optimum pixel size for a digital
detector.  For existing sites and all planned facilities we want to know: 1) the site
quality; 2) what kind of natural atmospheric variations there are on short time scales
and long time scales; and 3) whether population growth in the area is affecting the
astronomical site quality.  Another issue we will emphasize in this paper is that of
telescope baffling.  If a telescope is poorly baffled, then skylight is scattered
around the inside of the telescope, raising the background against which we are trying
to measure faint astronomical targets.  Not much can be done for old telescopes, but
this is a critical issue for the design and commissioning of new telescopes.

The literature on the subject of sky brightness is quite large and continues to grow.
In particular, the reader is directed to \citet{Roa_Gor73}, \citet{Lei_etal98}, and
references therein.  Data obtained at specific sites are described by \citet{Wal88},
\citet{Pil_etal89}, \citet[][hereafter K97]{Kri97}, \citet{Ben_Ell98}, \citet{Pat03},
and \citet{Pat07}.

\citet{Ray28} and \citet{Ray_Jon35} were the first to note a possible correlation 
between the sky brightness and the solar cycle.  See \citet{Wal88} for a more detailed 
discussion.  There are, of course, different measures of solar activity.  \citet[][Fig. 
4]{Wal88} shows a reasonably strong correlation between the $V$- and $B$-band sky brightness as a 
function of the 10.7-cm solar flux.\footnote[8]{The units of the solar flux are 
10$^{-22}$/m$^2$/Hz.  
For this paper we obtained the 10.7-cm solar flux values from
ftp://ftp.ngdc.noaa.gov/STP/SOLAR\_DATA/SOLAR\_RADIO/FLUX/.  These are the 
``Observed, Series C'' data from Penticton, British Columbia.} K97, Fig. 3, 
shows a good correlation of the zenith $V$-band sky brightness vs. the 10.7-cm solar 
flux if we eliminate data from the years 1985 and 1993 from the analysis.  It is not too 
surprising that there is a correlation of the $V$-band sky brightness and some measure 
of the solar activity.  The solar wind energizes the Earth's upper atmosphere, causing 
occasional auroral displays. A much smaller effect is the nightly airglow, which has the 
same origin.  The strong atmospheric emission line at 557.7 nm, attributed to [O I], 
falls in the $V$-band.\footnote[9]{We note that the Sloan Digital Sky Survey's 
$g^{\prime}$ and $r^{\prime}$ bands are strategically chosen so that the 557.7 nm line 
falls in between their response curves \citep{Fuk_etal96}.} While this line contributes 
directly to $V$-band sky brightness, the solar flux must also contribute to $B$-band sky 
brightness variations.  K97 found that the color of the sky was quite constant over the 
course of the solar cycle, with $\langle B-V \rangle= 0.930 \pm 0.018$.

In this paper we discuss 15 years of sky brightness measures obtained at Cerro Tololo
Inter-American Observatory.  The data were obtained with CCD detectors on the CTIO 0.9-m,
1.3-m, and 1.5-m telescopes.  The 0.9-m and 1.5-m telescopes were built in the late 1960's, so
are no longer modern telescopes.  The 1.3-m telescope was originally used for the Two Micron
All Sky Survey (2MASS).  Following the conclusion of that survey the 1.3-m was transferred to
CTIO, and, along with the 0.9-m, 1.0-m, and 1.5-m telescopes, became part of the Small and
Moderate Aperture Research Telescope System (SMARTS)  in 2003.  Preliminary analysis of the
CTIO data was discussed by \citet{Ver_etal02}.

\section{Deriving the Sky Brightness}

Our CCD-based data were reduced within the IRAF\footnote[10]{IRAF is distributed by
the National Optical Astronomy Observatory, which is operated by AURA, Inc.
under cooperative agreement with the National Science Foundation.} environment.
First the images are bias-corrected, trimmed, and flattened.  The 0.9-m images
are typically read out with four amplifiers, which have different effective gains,
but the resulting flattened frames show no significant background differences in
the four quadrants.  To calibrate the sky brightness data on any given night
we used 3 to 10 standards of \citet{Lan92}.  

Using {\sc dophot} or {\sc daophot} it is possible to determine the point spread
function (PSF) of the telescope and CCD camera for every frame, then, using this
information, subtract the stars, galaxies, and cosmic rays from the frames.  We did
{\em not} do this.  Instead, using some IRAF scripts written by one of us (NBS), we
simply made use of the {\sc imhist} program.  Since a majority of the pixels are
looking at sky, the mode of the pixel counts will correspond to the sky
level.\footnote[11]{We carried out tests with imagery of two globular clusters and
found that our IRAF scripts gave nearly identical sky brightness values compared to
those obtained using {\sc imstat} on small sub-areas or deriving the median sky counts
in a sky annulus while doing aperture photometry on more isolated stars at the edges of
the fields.} After iteratively clipping low and high pixels, we fit a Gaussian function
to the remainder of the data in the histogram.  The peak of this Gaussian fit gives us
the most robust value of the number of counts in the sky.  Of course, one assumes
that the master bias frame and overscan regions used for bias correction
remove the bias without the addition of any significant systematic effect. Any
problems with bias subtraction can be essentially eliminated by deriving the sky
brightness from frames having long exposures (e.g. 300 sec or longer).

Say we perform large aperture photometry on a standard star using {\sc apphot} within
IRAF, and this gives us a total of C$_{\star}$ counts above sky with an exposure time
of E$_{\star}$. The standard star is observed at airmass X$_{\star}$.  The atmospheric
extinction in that band (either assumed or derived) is $k_{\lambda}$.  Let the standard
magnitude of the star from \citet{Lan92} be M$_{\star}$.  Let C$_{sky}$ be equal to the
mean sky counts times the area of the software aperture in a {\em different} image with
exposure time E$_{sky}$.  Following Eq. 1 of K97, the magnitude of the sky signal is
then

\begin{equation}
S \; = \; -2.5 \; \rm{log} \; (C_{sky}/C_{\star}) + 2.5 \; \rm{log} \; 
(E_{sky}/E_{\star}) +   k_{\lambda} X_{\star} + M_{\star}  \; \; . 
\end{equation}

\parindent = 0 mm

One assumes that there are no systematic errors in the exposure times as given
by the data acquisition system.  Obviously, tests can and should be done to
investigate this question.  The basic rule is: longer exposures are better.

\parindent = 9 mm

Since the catalogue value of the standard star magnitude corresponds to its 
out-of-atmosphere value, one corrects the standard star signal for the extinction in the 
Earth's atmosphere by adding the term $k_{\lambda} X_{\star}$.  The sky brightness along 
some line of sight in the sky is {\em not} corrected to an out-of-atmosphere value. 
Given the plate scale of the CCD image (i.e., the number of arc seconds per pixel), we 
can calculate the area of the software aperture A, measured in square arc seconds. The 
sky brightness I($\mu$) in magnitudes per square arc second is then

\begin{equation}
I(\mu) \; = \; S \; + \; 2.5 \; \rm{log} \; A \; \; .
\end{equation}

Of course, one can also fit a PSF to the standard stars to obtain the number of counts
above sky.  The corresponding apparent magnitude of a sky patch can directly be
transformed into the sky brightness in magnitudes per square arc second by knowing the
plate scale and calculating the area of the sky patch.  Finally, one can use
measurements of multiple standard stars to give a more robust calibration of the sky
flux.  Since the sky has the color of a K0-type star, one should avoid blue standard
stars in order to eliminate as much as possible any filter effects.

Because magnitudes are a logarithmic system, for statistical purposes it is not correct
to average sky brightness values in mag/sec$^2$.  One should convert the data to some
kind of flux units, average them, then convert the numbers back to mag/sec$^2$.  
Following \citet{Gar89}, \citet{Sch90}, and K97, for the $V$- and $B$-bands one can use
nanoLamberts for the flux:

\begin{equation}
B_{obs}\rm{(nL)} \; = \; 0.263 \; a^{[Q - I(\mu)]} \; ,
\end{equation}

\parindent = 0 mm

where $a$ = (100)$^{0.2}$ $\approx$ 2.51189, Q = 10.0 + 2.5 log(3600$^2$)
$\approx$ 27.78151, I($\mu$) is the sky brightness in mag/sec$^2$, and the
factor 0.263 is the surface brightness (in nL) of a star with $V$ = 10 spread
out over one square degree.

\parindent = 9 mm

For airmass less than 1.6 (and possibly larger) it is appropriate to correct the 
observed sky brightness to the zenith value using Eq. 1 of \citet{Sch90}:

\begin{equation}
B_{zen} \; = \; B_{obs} / (1 \; + \; Z^2_{rad}/2) \; \; ,
\end{equation}

\parindent = 0 mm

where Z$_{rad}$ is the zenith angle in {\em radians}.

\parindent = 9 mm

As noted above, the CTIO 0.9-m and 1.5-m telescopes are 40 years old.  Our analysis
shows that the CTIO 1.3-m telescope gives, on average, demonstrably fainter sky
brightness values compared to data from the two much older telescopes.  After some
simple experiments in the dome, we attribute this to bad baffling in the older
telescopes.  The bottom line is that the camera window facing the Cassegrain
secondary mirror should only receive light from that secondary.  A poorly baffled
telescope will allow light scattering off the inside of a solid telescope tube to hit the
CCD camera window.  This will brighten the sky background.  A poorly baffled telescope 
with an open tube will allow light from the sky {\em and} light from the inside 
of the dome to degrade the measured sky brightness.

Using imagery obtained with the three telescopes during
2003, 2004, and 2005, we have derived baffle corrections for the data obtained with
the 0.9-m and 1.5-m telescopes (Table \ref{baffle}).\footnote[12]{This is to say 
that the baffle corrections are adjustments for systematic errors in the 0.9-m and 1.5-m
data. These adjustments could have systematic errors of their own, which we estimate
to be of order $\pm$~0.05 mag/sec$^2$.}  We assume explicitly that 
the more modern CTIO 1.3-m telescope is well baffled and the sky brightness values
from images obtained with it are correct.

Note that the baffle corrections increase monotonically with wavelength, reaching half
a magnitude in the $I$-band.  If there were other factors contributing to systematic
errors in our CTIO data from 1992 through 2002, it would be difficult to determine at
this stage.

Finally, we note that the $U$-band baffling corrections for the older telescopes
are inconsistent with the \bvri corrections, in the sense that they equal $-$0.27
mag/sec$^2$ for the 0.9-m and 0.00 mag/sec$^2$ for the 1.5-m.  If bad baffling is
the cause of the arithmetically positive corrections for the other filters, then
it does not make sense that the CCD camera on the 1.3-m would suffer local light
pollution only in the $U$ filter.

\section{A Sanity Check on Systematic Errors}

As a sanity check, we shipped to Chile the photometer and telescope used by K97 for his 
sky brightness measures obtained at Mauna Kea from 1985 through 1996 \citep{Kri96}.  
That system gives an elliptical footprint on the sky of 6.522 $\pm$ 0.184 square arc 
minutes and uses an RCA 931A photomultiplier tube.  Given the nature of this 
instrument, it was difficult to avoid stars fainter than V = 13 in the beam.  Ironically,
poor tracking allowed us to sample a small swath of sky and pick off the minimum sky signal.
We would expect that CCD-based sky brightness values would be somewhat fainter than 
data obtained with the Krisciunas system, since faint stars and galaxies can be eliminated 
from CCD analysis.

In Tables \ref{ccd_check} and \ref{pm_check} we give some sky brightness values obtained on
two photometric nights at CTIO in December of 2006.\footnote[13]{\citet{Lan92} fields were
observed in \bvri on 7 occasions over the course of 2006 December 23 and 24.  Using {\sc
evalfit} within the {\sc photcal} package, we found that the RMS uncertainties of the \bvri
magnitudes of the standards were between $\pm$~0.01 and $\pm$~0.02 mag on these nights.  
Extinction values were measured to $\pm$ 0.01 mag/airmass.  Thus, we judge these two nights to
be of excellent photometric quality.  For the calibration of the single channel photometer
data our principal standard stars were BS 1179 and $\zeta$ Cae.  Our check star was $\rho$
For. Their $B$ and $V$ magnitudes were obtained from \citet{Hof_Jas82}.} Table \ref{ccd_check}
gives data obtained with the CTIO 0.9-m telescope.  Some of the $V$-band sky brightness values
were obtained within two hours of the end of astronomical twilight (which occurred at roughly
01:16 UT on those nights).  The other CCD data were obtained at a fixed location on the sky,
RA = 5 hours, DEC = $-$30 degrees.  Table \ref{pm_check} gives data obtained with the
Krisciunas system at a number of positions west of the celestial meridian on the very same
nights.

Fig. \ref{dec23_24} shows the sky brightness measures obtained at CTIO with the two
different systems on 2006 December 23 and 24 UT.  Clearly, there is evidence that the
sky continued to get darker long after the nominal end of astronomical twilight. We
shall consider only the data obtained more than two hours after the end of astronomical
twilight.  In the $V$- and $B$-bands, respectively, the data from the Krisciunas system are,
on average, 0.13 and 0.17 mag/sec$^2$ brighter than the baffle-corrected 0.9-m data.  
These differences can be attributed to a combination of factors: 1) uncertainty in the
beam size of the Krisciunas system; 2)  the unknown contribution of faint stars in the 
Krisciunas system beam; and 3) systematic errors in the baffling corrections for the 0.9-m.  On the
whole, however, the data obtained with the Krisciunas system and the 0.9-m are in reasonable
agreement because one would expect the single channel photomultipler tube data to give
brighter values than CCD data based on pixels that were free of the light of stars and
galaxies.

We note that the recent data obtained with the Krisciunas system (corrected to the zenith)
give $\langle B-V \rangle = 0.906 \pm 0.034$, while the data from the 0.9-m obtained on
the same two nights (and more than two hours after the end of astronomical twilight)  
give $\langle B-V \rangle = 0.951 \pm 0.013$.  These values are in good agreement with
the average from K97 of $\langle B-V \rangle = 0.930 \pm 0.018$.

On 2006 December 23 and 24 we also measured the sky brightness at 10 to 11 degrees above the left
flank of La Serena.\footnote[14]{We could not aim directly over the center of the city because the
dome of the 0.9-m telescope was in the way.} In Figs. 4 and 5 of \citet{Gar89} we find the results of
his modelling the atmosphere at Boulder, Colorado (elevation 1655 m), and Mt. Graham, Arizona
(elevation 3267 m).\footnote[15]{A careful reading of the text of Garstang's paper reveals that the
captions to his Figs. 4 and 5 should be swapped.} Since CTIO is 2215 m above sea level, it makes sense
to average the two models for our purposes here.  We note, however, that the continental air of the
United States is not as aerosol-free as the maritime air of CTIO.  We assume that the total
contribution to the $V$-band sky brightness from directly transmitted light, Rayleigh scattering, and
aerosol scattering is 1.94 times brighter at a zenith angle of 79 or 80 degrees compared to the
contribution at the zenith.  For Garstang's Boulder model the value is 1.84 and for Mt. Graham the
value is 2.03.  In Table \ref{highZ_example} we convert some of our data from Table \ref{pm_check} to
fluxes in nL and compare the observed fluxes at high airmass with what we would predict on the basis
of the mean zenith sky brightness scaled by the factor from Garstang's model.

Luginbuhl (2007, private communication) indicates that on one recent occasion he and his
colleagues measured the sky brightness near Flagstaff, Arizona, to be $V$ = 21.85 mag/sec$^2$
at the zenith and 21.21 mag/sec$^2$ at elevation angle 10 degrees.  Those numbers translate
into a flux ratio of 1.80.  Whether for CTIO the most robust value of this parameter is 1.8 or 2.0,
we observed $\approx$ 3 times as much flux at high zenith angle compared to the zenith.

From the summit of Cerro Tololo one can look down at La Serena, Vicu\~{n}a, and
Andacollo and see artificial light with the naked eye if those locations are not
covered by cloud.  Statistically speaking, we obtained the {\em same} values of the sky
brightness at very high airmass on December 23 and 24.  At 10 to 11 degrees above the left
flank of La Serena we measured enhancements of 72 and 44 percent in the $V$-band on
the two nights in question. These are almost certainly measurements of light pollution
attributable to La Serena.  At elevation angles of 45 degrees or higher there is no
measurable effect on the night sky brightness at Tololo at this time.

\section{A Database of Useful CTIO Sky Brightness Measures}

Over the course of years of observing galaxies that have hosted supernovae,
we have accumulated many images.  These images can be used for the measurement of
the sky brightness at Cerro Tololo.  Of course, these images were taken under a
variety of sky conditions: photometric, non-photometric, with and without moonlight.
Some are short exposures.  Some are long exposures.  Some were taken during twilight
or when the zodiacal light was still strong. Some were taken in the middle of the night.  

Our database of images usable for measurement of the sky brightness involved an 
extensive selection process to reduce the effects of artifical brighteners of the sky.
This includes:

\parindent = 0 mm

1. Removal of images with exposure times shorter than 10 seconds.  Given the huge
number of pixels in a CCD chip, we find that it {\em is} possible to get
reliable sky brightness readings with exposures as short as 10 seconds.

2. Removal of images with airmass greater than 1.6.  The effect of dust and particles in
the Earth's atmosphere begins to dominate the sky brightness levels closer to the horizon.
See \citet[][Figs. 4, 5]{Gar89}.  Limiting the study to low airmasses reduces the 
effect of these particles on the sky brightness values.

3. Removal of images taken within 30 degrees of the Galactic plane.  Any image of the
night sky will contain countless unresolved sources which brighten the level of the sky.
By excluding images taken in the Galactic plane we significantly decrease the number of
unresolved stars that could contribute to this brightening.

4.  We include only images taken more than two hours after the end of evening
astronomical twilight (i.e. Sun 18 degrees below the horizon) until two hours before
the start of morning astronomical twilight. During astronomical twilight the sky is
being brightened by the Sun.  Up to two hours after the end of evening astronomical
twilight and starting two hours before the start of morning astronomical twilight the
sky is partially illuminated by the zodiacal light, which is sunlight scattering off
interplanetary dust.

5.  Removal of images taken when the Moon was above the horizon or if the Moon was
within 30 minutes of the horizon.

6.  Removal of images taken on non-photometric nights.  Any clouds would have a significant
impact on the observed brightness levels.  We consulted the historic sky conditions from the
CTIO website and excluded nights known to be non-photometric.

7. Removal of images more than three standard deviations from the mean on those nights
when multiple images were obtained.

8. For reasons outlined above, we choose to consider only $U$-band values obtained
with the CTIO 1.3-m telescope.

\parindent = 9 mm

In Table \ref{yearly_avgs} we give the yearly averages of the \bvri sky brightness
at CTIO.  Many of these yearly averages, especially during the 1990's, are based on a 
small number of images per year.  Of course, many other observers were using the 
CTIO 0.9-m and 1.5-m telescopes. We should have organized a system whereby 
observers could copy to disk deep images obtained in the middle of the night, along 
with images of standard stars.  The Paranal database described by \citet{Pat07} is
understandably more extensive than ours described here.  

As mentioned above, it is not correct to average data in magnitudes or mag/sec$^2$
because those are logarithmic units.  One should convert to fluxes, average the
fluxes, and then convert the average back to magnitude units if one so chooses.
This is what we have done in our analysis.


Fig. \ref{indiv} shows the individual zenith $V$-band sky brightness values derived
from CCD imagery obtained at CTIO.  While a solar cycle effect is apparent, we 
feel that yearly averages show the effect more clearly.

Fig. \ref{vsky} shows the yearly averages from K97 along with the CTIO yearly averages. There is
an overlap of four years.  As first reported by \citet{Ver_etal02}, the CTIO data of 1992 to
1994 are noticeably fainter than the data obtained at the 2800-m level of Mauna Kea and reported
by K97.  Even if we correct the Mauna Kea data of 1992 for the difference of solar flux levels
of the nights in question, we cannot reconcile the numbers.  The CTIO $V$-band data of 1992 are
based on 3 nights, so we could just be dealing with small number statistics.  Perhaps the baffle
corrections obtained from imagery of 2003 to 2005 are not the correct values to apply to the
data of 1992 through 1994.  The small amount of data obtained in 1996 at the two locations
matches within the errors, and the sanity check described in \S3 of this paper is reasonable
assurance that under careful conditions we get comparable values with the single channel system
and the CCD camera on the 0.9-m at CTIO.

In Fig. \ref{bri_sky} we show the yearly averages of the $BRI$ sky brightness at Mauna Kea
and at CTIO.  The Mauna Kea $B$-band data alone show a solar cycle effect, as does
the CTIO $B$-band dataset taken on its own.  However, as in the $V$-band, there is a serious
discrepancy as to zeropoint in the years 1992 to 1994.  We see no evidence for a solar cycle
effect in the $R$- and $I$-band data from CTIO.

Grand averages of CTIO and Paranal data are given in Table \ref{averages}.  The Paranal
data are based on images taken from April 2001 through April 2006 \citep{Pat07}.
Thus, both datasets cover years of solar maximum and solar minimum.  However, the years
2001 through 2006 are not equally represented in the Paranal data.  There are more observations
from 2001 to 2003 when the Sun was more active.  Patat (2007, private communication) indicates
that the long term $B$- and $V$-band sky brightness at Paranal is roughly 0.1 mag/sec$^2$
fainter than the values in Table \ref{averages}.

In Table \ref{averages} the uncertainties given are the standard deviations of the {\em
distributions}, not the standard deviations of the means.  Statistically speaking, the Paranal
data and the CTIO data are in agreement, given the typical standard deviations of $\pm$ 0.20
mag/sec$^2$.  With the 0.1 mag/sec$^2$ adjustment mentioned above, the CTIO data are, on
average, 0.06 mag/sec$^2$ fainter than Paranal in $B$ but {\em equal} in $V$.  This is evidence
that our baffling corrections are close to being correct, for these bands at least.

Under the reasonable assumption of a physical cause and effect between activity on
the Sun and the chemical reactions occurring in the Earth's atmosphere which result in
the airglow, we naturally
ask: is this due to the light which shines on the Earth eight minutes after leaving
the Sun's photosphere?  Or is it due to the solar wind, i.e. to particles coming
from the Sun?  

In Fig. \ref{vflux} we plot the yearly averages from Table \ref{yearly_avgs}, converted
to flux, vs. the mean of the 10.7-m solar flux 4.5 days prior to when the sky brightness was
measured.  We made various versions of this plot using solar flux values from the day
prior to a given night's observations until 8 days prior.  Since the solar flux is measured
about 0.5 d prior to a given night's observations, this corresponds to $-8.5 \leq  \Delta$ T 
$\leq -0.5$ d.  We find a minimum reduced $\chi^2$ value at $\Delta$T = $-$5.0 d.
Given the mean distance of the Earth from the Sun, a time delay of 5.0 days 
corresponds to a mean speed of the solar wind of $\approx$ 350 km s$^{-1}$.

This can be compared to the escape speed at the surface of the Sun, 618 km s$^{-1}$,
and to the speeds of the leading edges of coronal mass ejections, namely 
450 km s$^{-1}$ at solar maximum, and 160 km s$^{-1}$ at solar minimum
\citep{Kah00}.  More extensive photometry and sky spectra obtained at Paranal may
shed light on this time delay effect.\footnote[16]{From
http://solarscience.msfc.nasa.gov/SolarWind.shtml we can see a graph of the solar
wind velocity over the previous seven days.  A mean speed of 400 km s$^{-1}$ is
quoted, with a range of 300 to 800 km s$^{-1}$.}

As shown by \citet{Wal88}, \citet{Pil_etal89}, and K97, on any given night the sky brightness
can vary 10 to 50 percent.  There is not one {\em single} value for any given night.  
Whole-night wide-angle digital movies of the sky at CTIO obtained by Roger Smith show bands of
OH emission passing over the summit on time scales of tens of minutes.  It
is not surprising to measure variations of the airglow component of the sky brightness.

\section{Discussion}

Photometry of astronomical point sources in sparse fields is easy.  Photometry of
stars in crowded fields is more difficult.  Photometry of extended sources is much
more difficult because one must worry about seeing, contrast against the sky, and
plate scale.  Photometry of the night sky is of intermediate difficulty.  The
biggest systematic uncertainties arise from certain aspects of CCD observing that we
normally do not worry about: accuracy of exposure times, imperfect bias subtraction,
light leaks, and bad baffling in the telescopes.  

Ideally, one would like to be able to measure large solid angles of the sky and to
calibrate the observed sky brightness by means of many identifiable standard stars.
Such a system has been implemented, and is described by \citet{Dur_etal07}.  These
authors are able to image the entire sky over a span of half an hour and can obtain
robust photometric zeropoints and extinctions from the identification and detection of
over 100 bright standard stars in each dataset.

A comparison of sky brightness obtained with different equipment is largely a
search for systematic errors.  Because of the importance of northern Chile to
ground based observational astronomy, we felt it was important to calibrate the
night sky at Cerro Tololo using images easily available to us.  This also involved
taking data with the very same telescope and photometer used by \citet{Kri97}
for an 11 year study at Mauna Kea.  We find that observations obtained at
CTIO with the Krisciunas system are consistent with observations obtained with
the CTIO 0.9-m telescope if we adopt corrections for bad baffling in that telescope.

We have used an extensive database of images obtained for supernova research and have whittled
down the size of the database by excluding observations on non-photometric nights, observations
taken within two hours of the end or beginning of astronomical twilight, observations when the
Moon was within 30 minutes of the horizon, images obtained within 30 degrees of the Galactic
plane, and images taken at airmass greater than 1.6. The resulting database demonstrates a
correlation of the $V$-band sky brightness with the phase of the solar cycle, as has been
found by others over the past 80 years.  A solar cycle effect can be seen to a lesser extent
in the $B$-band data, but there appears to be no significant solar cycle effect in the $R$-
and $I$-band data.

We find that the $V$-band sky brightness is most tightly correlated with the solar
flux obtained 5 days prior to the night in question.  This corresponds to a mean
speed of $\approx$ 350 km s$^{-1}$ for the solar wind, in the mid-range of velocities
of coronal mass ejections at solar minimum and solar maximum.

We find no evidence of light pollution at Cerro Tololo within 45 degrees of the zenith
at this time.  However, 10 degrees over La Serena we measured a 58 $\pm$ 14 percent 
enchancement of the $V$-band sky brightness on two nights.  

\vspace {1 cm}

\acknowledgments

The CTIO 0.9-m, 1.3-m, and 1.5-m telescopes are operated by the Small
and Moderate Aperture Research Telescope System (SMARTS) Consortium.
DRS is grateful for the opportunity to participate in the Research
Experience for Undergraduates (REU) Program of the National Science Foundation.
JR thanks the Fulbright U.S. Student Program, IIE (The Institute of International
Education).  SV is grateful to the Pr\'{a}ctica de Investigac\'{i}­on en Astronom\'{i}a 
(PIA) Program.  We thank Ferdinando Patat for making data available ahead of publication,
and thank Chris Luginbuhl for useful discussions.
We particularly thank Malcolm Smith for his encouragement and support of this work.
This work was supported by Cerro Tololo Observatory.
An obituary of Hugo Schwarz will be published in an upcoming issue of the {\em Bulletin
of the American Astronomical Society}.  Other personal recollections can be found
at http://www.subjectivelens.com/Hugo/.


\newpage

\begin{deluxetable}{ccccc}
\tablewidth{0pc}
\tablecaption{Baffle Corrections\tablenotemark{a}\label{baffle}}
\tablehead{   \colhead{Telescope} &
\colhead{$B$} & \colhead{$V$} & \colhead{$R$} & \colhead{$I$} } 
\startdata
0.9-m & 0.269 & 0.274 & 0.369 & 0.521 \\
1.5-m & 0.132 & 0.286 & 0.322 & 0.527 \\
\enddata
\tablenotetext{a} {The values in the table are the number of magnitudes
per square arc second to {\em add} to the raw sky brightness data to
eliminate systematic differences in sky brightness compared to the
CTIO 1.3-m telescope.  These values are based on data taken during
 the years 2003 through 2005.}
\end{deluxetable}

\begin{deluxetable}{ccccccccc}
\tabletypesize{\scriptsize}
\tablewidth{0pc}
\tablecaption{Sky Brightness Values from CTIO 0.9-m Imagery\tablenotemark{a}\label{ccd_check}}
\tablehead{   \colhead{UT Date} & \colhead{$\langle$UT$\rangle$} & 
\colhead{RA} & \colhead{DEC} & 
\colhead{Filter} & \colhead{Exptime} & \colhead{Observed} & \colhead{Z} & 
\colhead{Corrected} } 
\startdata
Dec 23 & 01:07 & 23:36:39 &    $-$10$^{\rm o}$15$^{\prime}$ & $V$ & 300 & 21.262 & 44.83 & 21.552 \\
Dec 23 & 01:35 & 00:14:08 &    $-$10 25 & $V$ & 300 & 21.591 & 42.80 & 21.858 \\
Dec 23 & 02:06 & 02:08:18 & $-$3 50 & $V$ & 300 & 21.933 & 32.50 & 22.095 \\
Dec 23 & 03:00 & 02:20:38 & $-$7 54 & $V$ & 300 & 21.948 & 36.15 & 22.145 \\
Dec 23 & 04:08 & 05:00:00 & $-$30 00 & $V$ & 400 & 22.160 & \phs6.58 & 22.167 \\   
Dec 23 & 04:42 & 05:00:00 & $-$30 00 & $V$ & 400 & 22.171 & 15.90 & 22.212 \\   
Dec 23 & 05:05 & 05:00:00 & $-$30 00 & $V$ & 400 & 22.124 & 18.83 & 22.181 \\
\\ 
Dec 23 & 04:00 & 05:00:00 & $-$30 00 & $B$ & 600 & 23.081 & \phs4.91 & 23.085 \\
Dec 23 & 04:34 & 05:00:00 & $-$30 00 & $B$ & 600 & 23.072 & 12.23 & 23.096 \\
Dec 23 & 04:56 & 05:00:00 & $-$30 00 & $B$ & 600 & 23.053 & 17.16 & 23.101 \\
\\
Dec 24 & 01:05 & 23:29:44 & $-$9 37 & $V$ & 300 & 20.857 & 47.05 & 21.173 \\
Dec 24 & 01:31 & 00:28:38 &   +0 21 & $V$ & 300 & 21.241 & 46.75 & 21.553 \\
Dec 24 & 01:58 & 02:20:37 & $-$9 24 & $V$ & 300 & 21.742 & 25.58 & 21.845 \\
Dec 24 & 02:54 & 02:08:18 & $-$3 50 & $V$ & 300 & 21.503 & 40.92 & 21.750 \\
\\
Dec 24 & 03:53 & 05:00:00 & $-$30 00 & $V$ & 400 & 21.939 & \phs4.16 & 21.942 \\
Dec 24 & 04:13 & 05:00:00 & $-$30 00 & $V$ & 400 & 21.971 & \phs8.35 & 21.982 \\
Dec 24 & 04:32 & 05:00:00 & $-$30 00 & $V$ & 400 & 22.012 & 12.49 & 22.037 \\
Dec 24 & 04:53 & 05:00:00 & $-$30 00 & $V$ & 400 & 22.045 & 17.08 & 22.092 \\
Dec 24 & 05:11 & 05:00:00 & $-$30 00 & $V$ & 400 & 22.011 & 20.99 & 22.082 \\
Dec 24 & 05:30 & 05:00:00 & $-$30 00 & $V$ & 400 & 21.977 & 24.94 & 22.075 \\
Dec 24 & 05:49 & 05:00:00 & $-$30 00 & $V$ & 400 & 21.960 & 29.12 & 22.092 \\
Dec 24 & 06:07 & 05:00:00 & $-$30 00 & $V$ & 400 & 21.920 & 33.03 & 22.087 \\
\\
Dec 24 & 03:44 & 05:00:00 & $-$30 00 & $B$ & 600 & 22.991 & \phs2.49 & 22.992 \\
Dec 24 & 04:04 & 05:00:00 & $-$30 00 & $B$ & 600 & 22.898 & \phs6.68 & 22.905 \\
Dec 24 & 04:23 & 05:00:00 & $-$30 00 & $B$ & 600 & 22.952 & 10.81 & 22.990 \\
Dec 24 & 04:44 & 05:00:00 & $-$30 00 & $B$ & 600 & 22.978 & 15.42 & 23.017 \\
Dec 24 & 05:02 & 05:00:00 & $-$30 00 & $B$ & 600 & 22.991 & 19.32 & 23.051 \\
Dec 24 & 05:21 & 05:00:00 & $-$30 00 & $B$ & 600 & 22.968 & 23.28 & 23.054 \\
Dec 24 & 05:40 & 05:00:00 & $-$30 00 & $B$ & 600 & 22.937 & 27.46 & 23.055 \\
Dec 24 & 05:58 & 05:00:00 & $-$30 00 & $B$ & 600 & 22.916 & 31.38 & 23.068 \\
\enddata
\tablenotetext{a} {Year is 2006.  UT is in hours and minutes.
Right ascension is in hours, minutes, seconds (J2000).
Declination is in degrees and arc minutes. Exposure times are in seconds.  Column 7 is
observed sky brightness in mag/sec$^2$, using baffling corrections from
Table \ref{baffle}.  Z is the zenith angle in degrees.  Column 9 
data in mag/sec$^2$ are values from column 7, corrected to the zenith using
Eq. 4.}
\end{deluxetable}

\begin{deluxetable}{cccccccc}
\tablewidth{0pc}
\tablecaption{CTIO Sky Brightness Values from Single Channel System\tablenotemark{a}\label{pm_check}}
\tablehead{   \colhead{UT Date} & \colhead{$\langle$UT$\rangle$} & 
\colhead{RA} & \colhead{DEC} & 
\colhead{Filter} & \colhead{Observed} & \colhead{Z} & 
\colhead{Corrected} } 
\startdata
Dec 23 & 04:28 & 04:40 & $-$30 & $V$ & 22.038 (0.06) &  15.34 & 22.076 (0.06)  \\
Dec 23 & 04:47 & 01:50 & +18   & $V$ & 20.761 (0.04) &  78.54 & \nodata \\
Dec 23 & 05:15 & 05:00 & $-$30 & $V$ & 21.992 (0.06) &  21.18 & 22.064 (0.06)  \\
\\
Dec 23 & 04:38 & 04:40 & $-$30 & $B$ & 22.893 (0.10) &  17.50 & 22.943 (0.10)  \\
Dec 23 & 04:45 & 01:50 & +18   & $B$ & 22.266 (0.06) &  78.92 & \nodata \\
Dec 23 & 05:18 & 05:00 & $-$30 & $B$ & 22.884 (0.10) &  21.83 & 22.960 (0.10)  \\
\\
Dec 24 & 05:00 & 05:00 & $-$30 & $V$ & 21.747 (0.06) & 18.79 & 21.804 (0.06) \\
Dec 24 & 05:10 & 02:10 & +17   & $V$ & 20.782 (0.04) & 79.67 & \nodata \\
Dec 24 & 05:24 & 04:12 & \phs$-$5 & $V$ & 21.657 (0.06) & 45.11 & 21.950 (0.06) \\
Dec 24 & 05:47 & 05:00 & $-$30 & $V$ & 21.801 (0.06) & 28.91 & 21.951 (0.06) \\
\\
Dec 24 & 05:00 & 05:00 & $-$30 & $B$ & 22.750 (0.10) & 18.79 & 22.807 (0.10) \\
Dec 24 & 05:12 & 02:10 & +17   & $B$ & 22.290 (0.06) & 80.06 & \nodata \\
Dec 24 & 05:22 & 04:12 & \phs$-$5 & $B$ & 22.618 (0.10) & 44.72 & 22.907 (0.10) \\
Dec 24 & 05:44 & 05:00 & $-$30 & $B$ & 22.634 (0.10) & 28.27 &  22.759 (0.10) \\

\enddata
\tablenotetext{a} {Year is 2006.  UT and right ascension are in hours and minutes.
Declination is in degrees.   Column 6 is observed sky brightness in mag/sec$^2$.  
Z is the zenith angle in degrees.  Column 8 data in mag/sec$^2$
are values from column 6, corrected to the zenith using Eq. 4. Values in parentheses
are estimated random errors.}
\end{deluxetable}

\begin{deluxetable}{ccccccc}
\tablewidth{0pc}
\tablecaption{Detection of Artificial Light
at High Zenith Angle\tablenotemark{a}\label{highZ_example}}
\tablehead{   \colhead{UT Date} & \colhead{Filter} & \colhead{$B_{zen}$(nL)} & 
\colhead{$B_{obs}$(nL)} & \colhead{Ratio(obs/zen)} &
\colhead{$B_{pred}$(nL)} & \colhead{Ratio(obs/pred)} }
\startdata
Dec 23 & $V$ &  50.7 & 169.1 & 3.34 & 98.4 & 1.72 \\
Dec 23 & $B$ &  22.5 & \phs42.3 & 1.88 & \nodata  & \nodata \\
Dec 24 & $V$ &  59.6 & 165.9 & 2.78 & 115.6 & 1.44 \\
Dec 24 & $B$ &  25.3 & \phs41.4 & 1.64 & \nodata  & \nodata \\
\enddata
\tablenotetext{a} {Year is 2006. The values in column 6 are equal to the values in
column 3 times 1.94.  This scaling factor is obtained from averaging models of
one lower elevation site and one higher elevation site from Figs. 4 and 5 of
\citet{Gar89} and corresponds to a zenith angle of 79 to 80 degrees.}
\end{deluxetable}

\begin{deluxetable}{ccccccccc}
\tablewidth{0pc}
\tablecaption{Yearly Averages of Sky Brightness at CTIO\tablenotemark{a}\label{yearly_avgs}}
\tablehead{   \colhead{Year} & 
\colhead{$\langle B \rangle$} & \colhead{N$_B$} &
\colhead{$\langle V \rangle$} & \colhead{N$_V$} &
\colhead{$\langle R \rangle$} & \colhead{N$_R$} &
\colhead{$\langle I \rangle$} & \colhead{N$_I$} }
\startdata
1992 & 22.971 (0.024) & 3  & 21.842 (0.056) & 5  & \nodata        & \nodata & \nodata & \nodata \\
1993 & 23.122 (0.085) & 2  & 21.897         & 1  & \nodata        & \nodata & \nodata & \nodata \\
1994 & 23.259 (0.022) & 5  & 22.034 (0.012) & 4  & \nodata        & \nodata & \nodata & \nodata \\
1996 & 22.964 (0.050) & 2  & 21.904 (0.005) & 2  & \nodata        & \nodata & 19.956         & 1 \\
1997 & 22.745 (0.057) & 11 & 21.803 (0.051) & 15 & \nodata        & \nodata & \nodata & \nodata \\
1998 & 22.982 (0.074) & 3  & 21.911 (0.018) & 4  & \nodata        & \nodata & \nodata & \nodata \\
1999 & 22.741 (0.018) & 13 & 21.600 (0.039) & 11 & \nodata        & \nodata & \nodata & \nodata \\
2000 & 22.766 (0.048) & 8  & 21.564 (0.055) & 11 & 20.880         & 1  & 19.374         & 1  \\
2001 & 22.870 (0.024) & 11 & 21.668 (0.052) & 15 & 21.110 (0.190) & 3  & 19.828(0.172)  & 5  \\
2002 & 22.676 (0.067) & 10 & 21.694 (0.045) & 13 & 21.162 (0.067) & 13 & 19.895 (0.061) & 18 \\
2003 & 22.815 (0.028) & 42 & 21.817 (0.020) & 78 & 21.208 (0.018) & 65 & 19.814 (0.032) & 71 \\
2004 & 22.772 (0.025) & 49 & 21.710 (0.024) & 63 & 21.085 (0.019) & 60 & 19.848 (0.035) & 57 \\
2005 & 22.834 (0.026) & 38 & 21.854 (0.024) & 95 & 21.278 (0.022) & 83 & 19.866 (0.021) & 95 \\
2006 & 22.994 (0.032) & 13 & 22.061 (0.031) & 12 & 21.018 (0.003) & 2  & 19.726 (0.032) & 2  \\
\enddata
\tablenotetext{a} {Values are measured in mag/sec$^2$.
The numbers in parentheses are 1-$\sigma$ uncertainties (mean
errors of the mean).  There are no data from 1995. N$_i$ is the number of images,
not the number of nights.}
\end{deluxetable}

\begin{deluxetable}{ccccccccccc}
\tabletypesize{\scriptsize}
\tablewidth{0pc}
\tablecaption{Mean Sky Brightness at CTIO and Paranal\tablenotemark{a}\label{averages}}
\tablehead{  \colhead{Site} & 
\colhead{$\langle U \rangle$} & \colhead{N$_U$} & 
\colhead{$\langle B \rangle$} & \colhead{N$_B$} &
\colhead{$\langle V \rangle$} & \colhead{N$_V$} &
\colhead{$\langle R \rangle$} & \colhead{N$_R$} &
\colhead{$\langle I \rangle$} & \colhead{N$_I$} }
\startdata
CTIO       & 22.12 (0.19) &  27 & 22.82 (0.19) &  210 & 21.79 (0.22) &  329 & 21.19 (0.19) &  227 & 19.85 (0.25) &   250 \\
Paranal    & 22.35 (0.19) & 261 & 22.66 (0.16) & 1332 & 21.69 (0.21) & 1619 & 20.91 (0.23) & 3595 & 19.65 (0.28) &  2882 \\
difference & $-$0.23     & \nodata &  0.16 & \nodata        & 0.10         & \nodata & 0.28 &  \nodata      & 0.20      & \nodata \\
\enddata 
\tablenotetext{a} {Sky brightness is measured in mag/sec$^2$. $U$-band average from CTIO
is from images taken with the 1.3-m telescope only.  Paranal values from \citet{Pat07}
are based on data from April 2001 through April 2006, but the Paranal averages are
weighted more toward 2001 to 2003, when the solar cycle was closer to maximum.
The values in parentheses are the standard deviations of the
{\em distributions}, not the standard deviations of the means.}
\end{deluxetable}

\clearpage

\figcaption[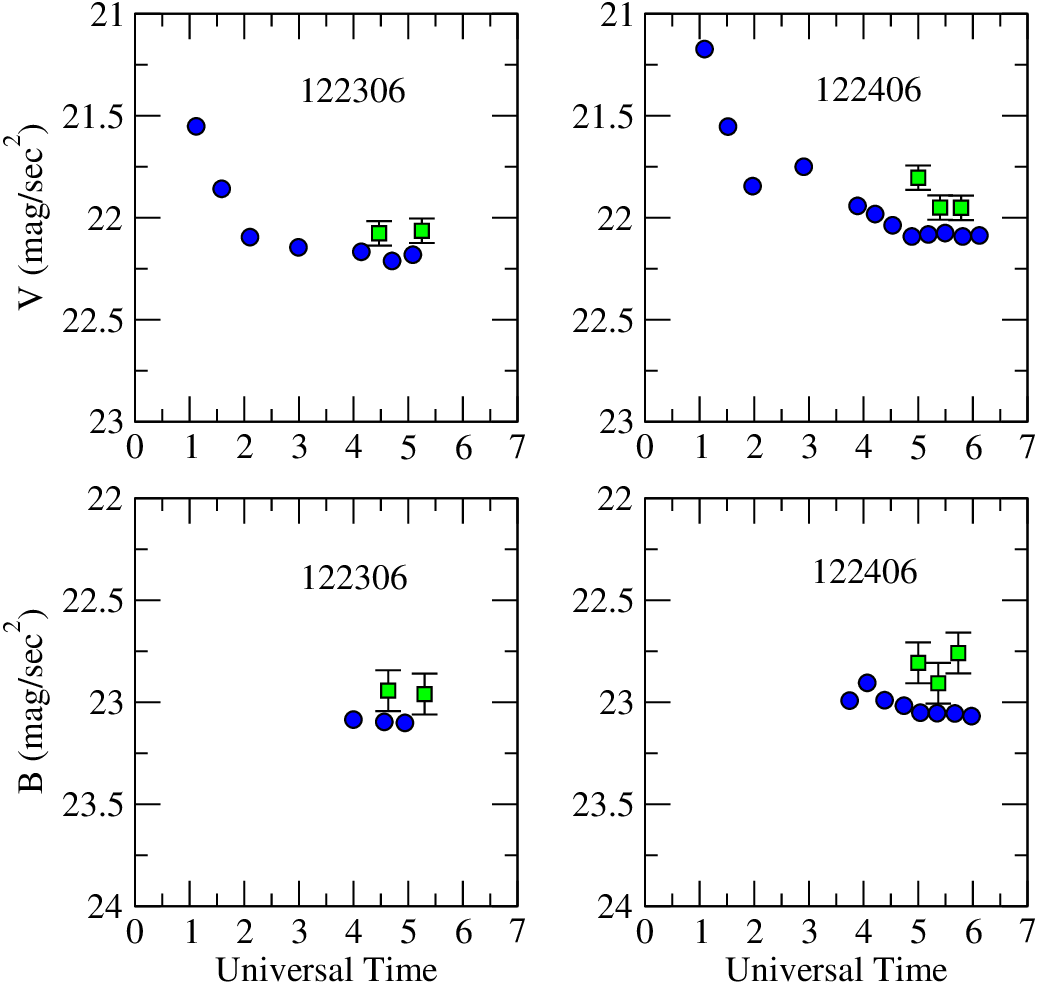]
{Values of sky brightness from CTIO.  Top panels: $V$-band sky brightness.
Bottom panels: $B$-band sky brightness.  Left panels: data of 2006 December
23 UT.  Right panels: data of 2006 December 24 UT.  The blue dots are data
from the CTIO 0.9-m reflector.  The green squares are data obtained with the
single channel system of \citet{Kri96}.  All data have been corrected to the
zenith. \label{dec23_24}
}

\figcaption[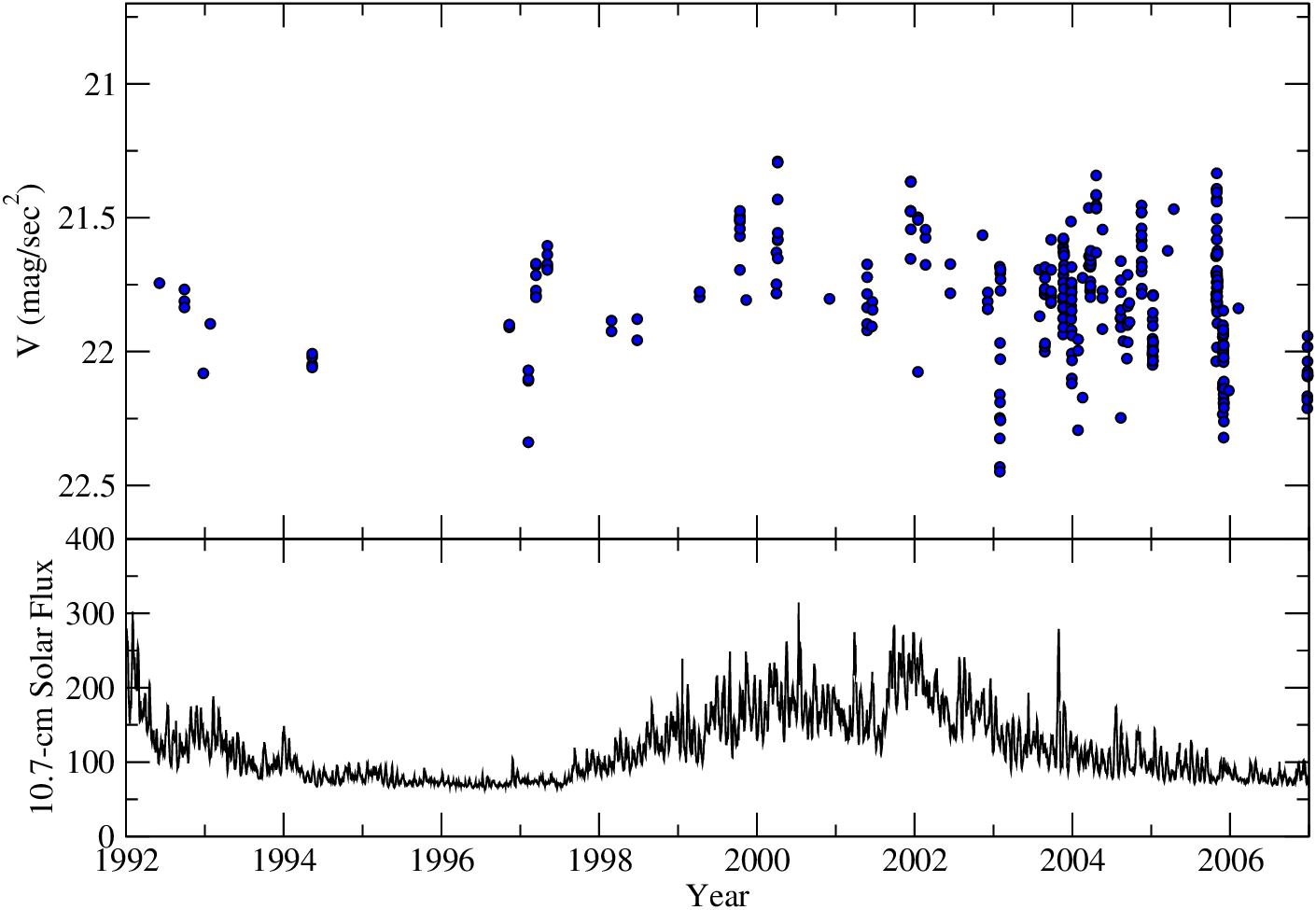]
{Upper panel: Individual CCD-based values of zenith $V$-band sky brightness from 
CTIO.  Lower panel: 10.7-cm solar flux. \label{indiv}
}

\figcaption[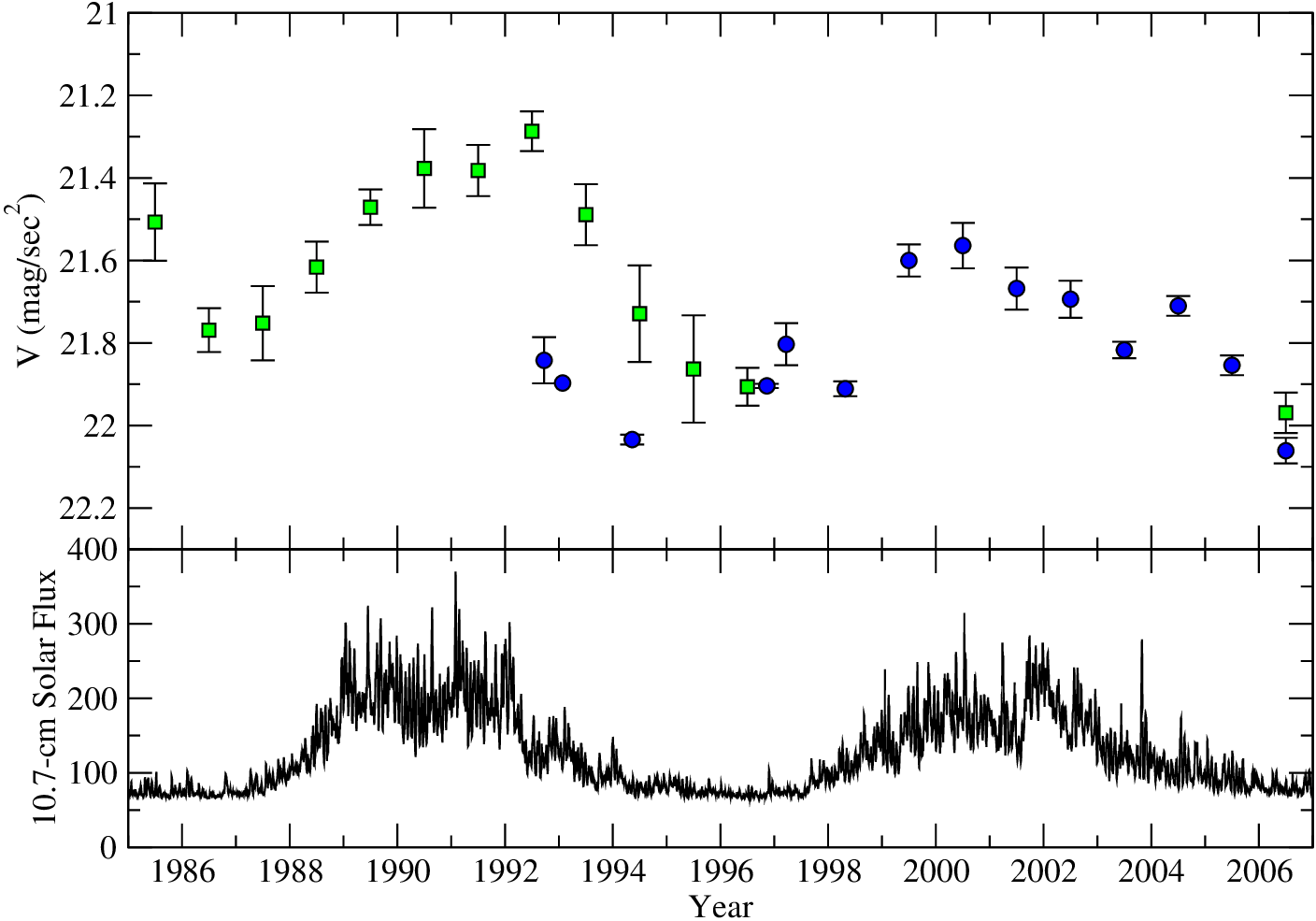] {Upper panel: Data obtained at the 2800-m level of Mauna Kea using a
15-cm telescope and single channel photometer \citep{Kri96}, along with
the average of the data obtained at CTIO in December of 2006 (green squares).  
These data were typically taken within 20 degrees of the zenith.
The blue dots are yearly averages of data obtained at CTIO using CCD imagery.
The CTIO data and the single channel data from December 2006
have all been reduced to the zenith.  Lower panel: 10.7-cm solar flux.
\label{vsky}
}
\figcaption[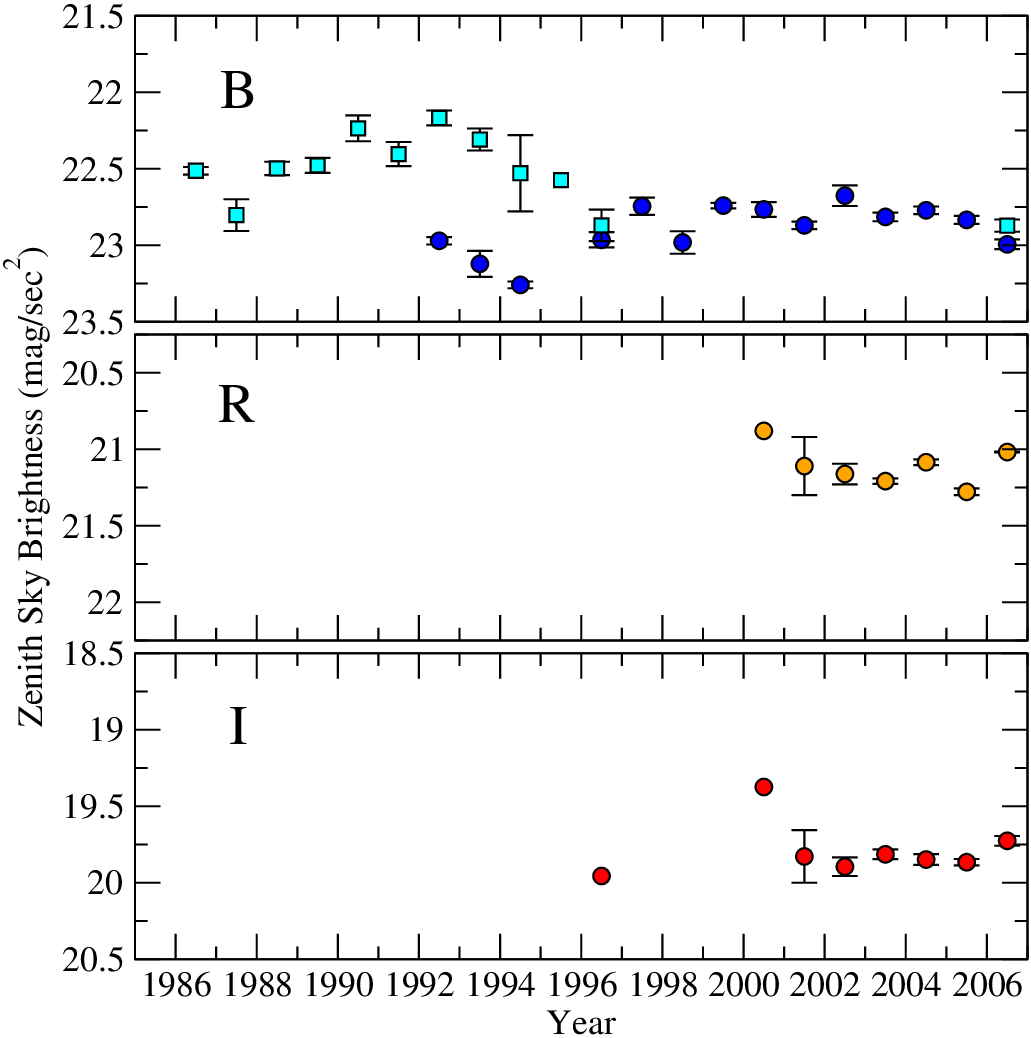] {Yearly averages of $BRI$ sky brightness.  In the top panel
the squares represent data from K97, along with the data from Table \ref{pm_check} of this
paper.  Circles represent CCD-based data from CTIO presented in this paper.
\label{bri_sky}
}

\figcaption[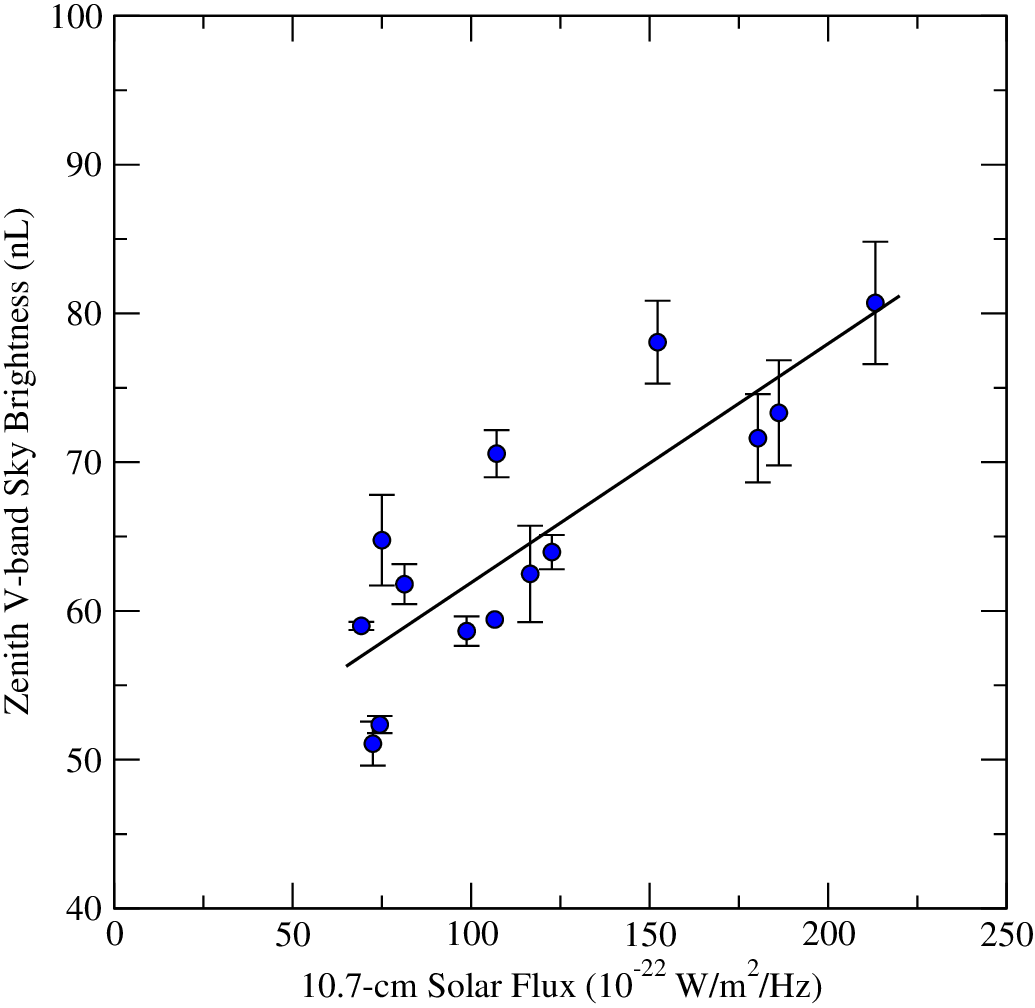] {Yearly averages of zenith $V$-band sky brightness
obtained from CCD imagery at Cerro Tololo (converted to flux) vs. the average of the
10.7-cm solar flux 4.5 days prior to when the sky brightness was measured.  
\label{vflux}
}

\clearpage

\begin{figure}
\plotone{dec23_24.eps}
{\center Krisciunas {\it et al.} Fig. \ref{dec23_24}}
\end{figure}

\begin{figure}
\plotone{indiv.eps}
{\center Krisciunas {\it et al.} Fig. \ref{indiv}}
\end{figure}

\begin{figure}
\plotone{vsky.eps}
{\center Krisciunas {\it et al.} Fig. \ref{vsky}}
\end{figure}

\begin{figure}
\plotone{bri_sky.eps}
{\center Krisciunas {\it et al.} Fig. \ref{bri_sky}}
\end{figure}

\begin{figure}
\plotone{vflux4.eps}
{\center Krisciunas {\it et al.} Fig. \ref{vflux}}
\end{figure}

\end{document}